\documentclass[twocolumn,secnumarabic,amssymb, nobibnotes, aps, prd]{revtex4}

\usepackage{color}
\usepackage{pstricks}
\usepackage{graphicx}

\begin{document}

\title{Domain walls in gapped graphene}%

\author{G.~W.~Semenoff, V.~Semenoff and Fei Zhou}%
\affiliation{Department of Physics and Astronomy, University of
British Columbia\\6224 Agricultural Road, Vancouver, British
Columbia V6T 1Z1}
\begin{abstract}The electronic properties of
a particular class of domain walls in gapped graphene are investigated.
We show that they can support mid-gap states which are localized in
the vicinity of the domain wall and propagate along its length.  With
a finite density of domain walls, these states can alter the
electronic properties of gapped graphene significantly.  If the
mid-gap band is partially filled,the domain wall can behave like a
one-dimensional metal embedded in a semi-conductor, and could
potentially be used as a single-channel quantum wire.
\end{abstract}
\maketitle

Graphene is a one-atom thick layer of carbon atoms with a hexagonal
lattice structure and where electrons within $\sim1ev$ of the Fermi
energy obey a Dirac equation and have a linear dispersion relation
$\omega =v_F|\vec k|$ with Fermi velocity $v_F\sim c/300$. It has been
studied as an analog of relativistic field theory
\cite{Semenoff:1984dq} where relativistic quantum mechanics and field
theory phenomena special to 2+1 space-time dimensions
\cite{Niemi}\cite{jackiw} could be realized in nature.  It was
identified and studied in the laboratory in 2004 \cite{novoselov:2004}
and it turns out to have high conductivity and carrier mobility and
other interesting properties which make it a promising material for
applications in electronic devices
\cite{Novoselov:2005kj}-\cite{kats}.

An important current problem is to modify graphene so that it has a
gap in its energy spectrum \cite{mindthegap}.  The idea is that, with
a small gap, so that electrons obey a Dirac equation with mass, it
would retain the good features of graphene and could also be used as a
semi-conductor in applications where a gap is essential, for example,
a field effect transistor. Breaking the symmetry which interchanges
the two triangular sublattices of the hexagonal graphene lattice will
gap the spectrum \cite{Semenoff:1984dq}.  This could be accomplished,
for example, by giving electrons residing on $A$ sites a different
energy from those on $B$ sites by introducing a staggered chemical
potential \cite{Semenoff:1984dq}.  It can also arise from deformations
of bonds on the graphene lattice \cite{chamon:2007fr} analogous to
those known from the study of carbon nanotubes \cite{nanotubes}.  A
third possibility is to use multi-layer graphene where the layers can
be stacked so that their interaction breaks the sublattice
symmetry. In all cases, to retain the features of the Dirac equation
the gap should be much less than the nearest neighbor hopping amplitude $t
\sim 2.7ev$.

The diatomic material Boron-Nitride (BN) has the same lattice
structure and valence electrons as graphene and a staggered chemical
potential by virtue of having different atoms on the two sublattices.
Monolayers have been made in the laboratory \cite{Novoselov:PNAS}.
However, the gap is too large $\sim 4.5ev$ for Dirac electrons. An
approach currently being pursued is to attach a graphene monolayer to
a BN substrate.  The resulting gap in graphene is estimated to be
$\sim 53mev$ \cite{bn} which is in the interesting range. Another
approach is epitaxial growth of graphene on a Silicon-Carbide
substrate where a larger magnitude gap $\sim2.6ev$ has been observed
\cite{lanzara}.

In this Paper, we shall consider line-like domain wall defects in the
mass pattern in graphene which is gapped by a sublattice symmetry
breaking staggered chemical potential. We shall find that they can
have significant electronic properties.  The domain walls are shown in
Figs.~1(b),1(c) where the zig-zag and armchair walls form boundaries
between regions where the staggered chemical potential is shifted
between the two sublattices. Such domain walls could be realized
naturally in BN and would be inherited by graphene on a BN substrate,
for example.  We shall show that they can give rise to a band of
mid-gap states. These states are localized in the vicinity of the wall
and propagate along its length.  If the mid-gap band is partially
filled, the domain wall can behave like a one-dimensional metal
embedded in a semi-conductor, and could potentially be used as a
single-channel quantum wire. One might imagine that, once techniques
for deposition of graphene monolayers on substrates are better
developed, the conditions for existence of these domain wall wires
could be created and manipulated to the point where they could be used
to print electric circuits on graphene sheets.

Mid-gap states already play an important role in graphene. It was
pointed out long ago \cite{Semenoff:1984dq} that an index theorem
governs the degeneracy of the $E=0$ Landau level in graphene in a
magnetic field and this level is half-filled in the neutral material.
This observation has spectacular experimental confirmation in the
half-odd-integer anomalous quantum Hall effect
\cite{Novoselov:2005kj}\cite{halleffect}-\cite{Gusynin:2005pk}. In
addition, theoretical studies of point-like vortex defects in a mass
condensate due to a Kekule distortion of graphene find mid-gap
electron states which can give the vortices fractional charge
~\cite{chamon:2007fr}\cite{JackiwPi}-\cite{Pachos}, thus giving a two
dimensional realization of a phenomenon previously known to occur in
one-dimensional linearly conjugated polymers such as polyacetylene
\cite{polyacetylene1}-\cite{polyacetylene3}.  Similar states bound to
vortices in a proximity-induced superconducting condensate in graphene
could lead to anyonic statistics with potential applications to
quantum computing~\cite{wilczek}. An essential common feature of these
examples is the existence of ``zero-mode'' mid-gap states in the
spectrum of the Dirac Hamiltonian which arise from the interaction
with fields that have a non-trivial topology. In the case of the
vortex, this topology is due to the vorticity.  Let us consider a
simple example to show that a related phenomenon takes place for a
domain-wall. Consider the $4\times4$-matrix graphene Dirac Hamiltonian
with the addition of a mass term (which might arise from a staggered
chemical potential):
\begin{equation}\label{dirachamiltonian}
H=\hbar v_F\left[ \matrix{ \frac{mv_F}{\hbar}  &
i\frac{d}{dx}+\frac{d}{dy} & 0 & 0 \cr i\frac{d}{dx}-\frac{d}{dy}
&-\frac{mv_F}{\hbar}  &0&0\cr 0&0&\frac{mv_F}{\hbar} &
i\frac{d}{dx}-\frac{d}{dy} \cr 0 & 0 & i\frac{d}{dx}+\frac{d}{dy}&
 -\frac{mv_F}{\hbar} \cr}  \right]
\end{equation}
The two diagonal blocks correspond to the two graphene valleys,
which transform into each other under parity and time reversal.
A domain wall is described by replacing the mass $m$ in
Eq.~(\ref{dirachamiltonian}) by a function $m(x)$ which depends on
one of the coordinates, $x$ with a soliton profile
\begin{equation}\label{solitonasymptotics}
\lim_{x\to-\infty}m(x)=-m <0 ~~,~~ \lim_{x\to\infty}m(x)=m
>0
\end{equation}
The energy spectrum then has the same gapped conduction and valence
band branches as would occur if $m(x)$ in Eq.~(\ref{dirachamiltonian})
were a constant with the asymptotic value of the mass, $m$: $E = \pm
v_F\sqrt{\vec k^2 +  m^2v_F^2 }$. These describe electrons in the
bulk semi-conductor away from the wall. As well, there is a gapless
mid-gap branch whose wave-functions have support near the wall.
Explicitly, the (un-normalized) wave-functions and eigenvalues are
\begin{eqnarray}
\psi_L(x,y) = e^{ik_yy/\hbar-\frac{v_F}{\hbar^2}\int_0^x dx'm(x')}
\left[ \matrix{ i\cr 1\cr 0\cr 0 \cr }\right] ~~E =  v_Fk_y
 \label{leftmover}\\  \psi_R(x,y) =
e^{ik_yy/\hbar-\frac{v_F}{\hbar^2}\int_0^x dx'm(x')} \left[ \matrix{
0\cr 0\cr 1\cr i \cr }\right] ~~E=-  v_Fk_y\label{rightmover}
\end{eqnarray}
Note that this solution exists and is continuum normalizable for
whatever the profile of the position-dependent mass term, it only
needs to have the asymptotic behavior of a topological soliton
as in Eq.~(\ref{solitonasymptotics}) \cite{footnote}. In particular, it should
be applicable to a one lattice spacing thick domain wall such as those
drawn in Figs.1(b),1(c).

What we have found are two bands of mid-gap states corresponding to
one left- and one right-moving one-dimensional massless fermion (for
each spin degree of freedom) traveling along the length of the domain
wall. An effective Lagrangian describing them would be
\begin{eqnarray}\label{effectiveaction}
L=i\sum_s\left[
\psi_{Ls}^\dagger(\partial_t-v_F\partial_x)\psi_{Ls} +
\psi_{Rs}^\dagger(\partial_t+v_F\partial_x)\psi_{Rs}\right]
 \end{eqnarray}
where $s$ labels the two spin states. Effects of impurities and local
interactions can be important in one dimension and should be taken
into account. Four-fermion operators are perturbatively marginal and
adding those which do not implement umklapp processes yields the
Tomonaga-Luttinger model which is a solvable conformal field theory
with well-known properties.

\begin{figure}
\begin{center}
\includegraphics[scale=0.6]{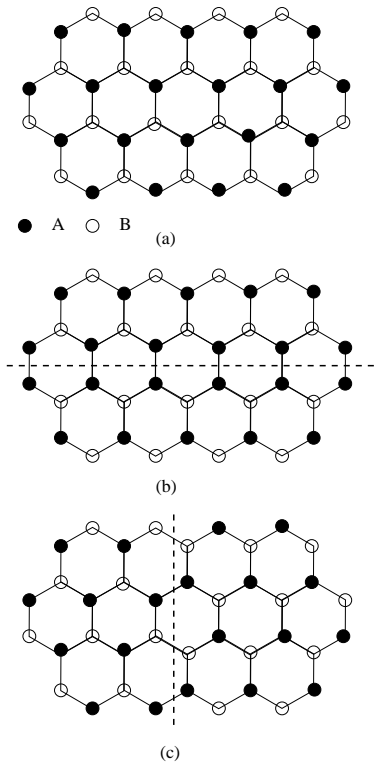}
\caption{ a)A hexagonal graphene lattice with triangular
sublattices $A$ (black dots) and $B$ (white dots) connected by vectors
$\vec s_1=\left(0,-1\right)a,\vec
s_2=\left(\sqrt{3}/{2},1/2\right)a,\vec
s_3=\left(-\sqrt{3}/2,1/2\right)a$ with lattice
constant $a=1.23{\AA}$. b)
A zig-zag domain wall.
c) An armchair domain wall.
In b) and c), the
chemical potentials are $-\mu$ on black dots
and $+\mu$ on white dots; the sublattices which have higher/lower
chemical potentials are interchanged at a domain wall
creating a line of miss-matched neighbors, denoted by
a dashed line.
}
\end{center}
\label{fig1}
\end{figure}

To understand the structure of the bands in more detail, we must take
a closer look at the tight-binding lattice model of gapped
graphene. We shall find features which are not reflected in the
continuum analysis which is only valid in a small region near
$E=0$. The Hamiltonian is
\begin{eqnarray}\label{tightbinding}
H=\sum_{A,i} t~b_{A+\vec s_i}^\dagger a_A +
\sum_{B,i}t~a^\dagger_{B-\vec s_i}b_{B}\nonumber \\
+ \sum_A\mu a_A^\dagger a_A -\sum_B\mu b_B^\dagger b_B
\end{eqnarray}
where $a_A^\dagger,a_A,b_B^{\dagger},b_B$ are the quantum amplitudes
for an electron to occupy sites labeled $A$ and $B$ on the sublattices
$A$ and $B$, respectively. The lattice and sublattices are depicted in
Fig.~1(a). The first terms in Eq.~(\ref{tightbinding}) describe electron
tunneling between nearest neighbor sites. The terms proportional to
$\mu$ are on-site energies.  They break the sublattice symmetry and
generate a gap.
The Schr\"odinger equation is
\begin{eqnarray}\label{schrodinger}
(E-\mu)a_A = t\sum_i b_{A+\vec s_i}~~,~~ (E+\mu)b_B = t\sum_i
a_{B-\vec s_i}
\end{eqnarray}
To study the zig-zag domain wall in Fig.~1(b), we solve
Eq.~(\ref{schrodinger}) with $\mu$ replaced by $\mu~{\rm sign}(A_y)$
and $\mu~{\rm sign}~(B_y)$.  The spectrum has branches corresponding
to electrons propagating in the bulk of the gapped graphene away from
the wall, {\small \begin{equation}\label{bulk} E = \pm \sqrt{ \mu^2
+t^2\left( 2\cos^2\frac{\sqrt{3}a}{2}k_x +
\cos\frac{3a}{2}k_y\right)^2+t^2\sin^2\frac{3a}{2}k_y}
\end{equation}}
Here, $(k_x,k_y)$ are wave-vectors.
This bulk spectrum has a gap $2\mu$ and is symmetric about $E=0$.
Then, there are two branches with wave-functions which
fall off exponentially with transverse distance $|A_y|,|B_y|$ from the
wall and are oscillating functions of the longitudinal
$A_x,B_x$-coordinates with wave-vector $k_x$:
\begin{eqnarray}\label{zigzag1}
E =-t-\sqrt{ \mu^2+4t^2\cos^2\frac{\sqrt{3}a}{2}k_x}
~,~\cos\frac{\sqrt{3}a}{2}k_x\leq0
\\ \label{zigzag2}
E=t-\sqrt{ \mu^2+4t^2\cos^2\frac{\sqrt{3}a}{2}k_x} ~,~
\cos\frac{\sqrt{3}a}{2}k_x\geq 0
\end{eqnarray}
The band in Eq.~(\ref{zigzag1}) is located inside (and reaches
slightly below) the negative energy bulk states Eq.~(\ref{bulk}). The
other branch Eq.~(\ref{zigzag2}) covers the interval
$[t-\sqrt{\mu^2+4t^2},t-\mu]$.  If $t>\mu$, this band crosses zero
energy ($E=0$) at two values of $k_x$, and thus agrees with the
continuum analysis which is only valid when $\mu<<t$ and which
predicts the existence of two zero energy modes -- one for each
crossing. The spectrum and density of states for $\mu=0.5t$ are
depicted in Fig.~2.  Note that, unlike the spectrum of bulk graphene
Eq.~(\ref{bulk}) the zig-zag domain wall spectrum is not symmetric
about $E=0$.  This is evident from its structure displayed in
Fig.~1(b), where the miss-matched sites along the wall are entirely
black dots with chemical potential $-\mu$.  The zig-zag domain wall
violates the symmetry which reflects the sign of the energy.  There is
an anti-wall where the miss-matched bonds are entirely white dots -
with energy $+\mu$. Its domain wall spectrum would have opposite sign
to Eqs.~(\ref{zigzag1}) and (\ref{zigzag2}).

We can get an intuitive understanding of the spectrum in
Eqs.~(\ref{zigzag1}) and (\ref{zigzag2}) in the limit where $\mu$ is
large.  Initially, neglecting $t$, there are two energy levels, $\mu$
for an electron sitting on a white dot and $-\mu$ for an electron
sitting on a black dotin Fig.~1(b).  Then, if we turn on small $t$,
the largest effect is for the black dots on each side of the domain
wall which have a nearest neighbor at the same zeroth order energy,
$-\mu$.  Turning on the hopping would split the degeneracy of these
sites to $-\mu+t$ and $-\mu-t$. Note that this does not happen for
sites in the bulk away from the domain wall, since they are not
degenerate with their neighbors -- corrections to their spectrum would
be at the next higher order in $t$.  The energies $-\mu- t$ and
$-\mu+t$ are identical to the Taylor expansions of
Eqs.~(\ref{zigzag1}) and (\ref{zigzag2}), respectively, to first order
in t. The next order in the hopping amplitide, second order
perturbation theory, would take into account hopping to an adjacent
site with energy $+\mu$ and back and would be of order ${t^2}/{\mu}$,
also what one would expect from expanding Eqs.~(\ref{zigzag1}) and
(\ref{zigzag2}) as well as (\ref{bulk}) to second order in t. The
order $t^2/\mu$ contributions are momentum dependent and the energy
levels become bands.  Then $t$ is made larger than $\mu$, they spread
out into the bands depicted in Fig.~2.

\begin{figure}
\begin{center}
\includegraphics[width=\columnwidth]{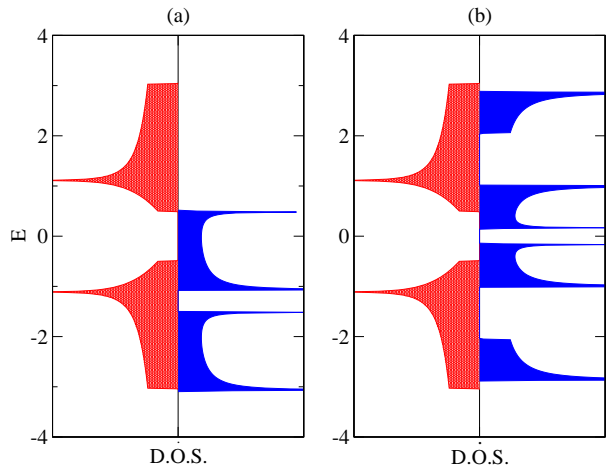}
\caption{(color online) Density of states (DoS in arbitrary units, 
colored blue) versus
energy $E$ (in units of t)
for a zig-zag domain
wall a) and an armchair wall b) when $\mu=0.5 t$.
Also shown is the DoS for bulk bands (red).}
\end{center}
\end{figure}

In the neutral ground state, the half of the electron states with
lowest energy will be filled. For the zig-zag wall there is a profound
difference between two cases - when the midgap band Eq.~(\ref{zigzag2})
overlaps the negative energy bulk band (${\mu}/{t}<{3}/{2}$)
and when it doesn't (${\mu}/{t}>{3}/{2}$).  When it doesn't,
the neutral ground state has the negative energy bulk states
Eq.~(\ref{bulk}) and the lower domain wall band Eq.~(\ref{zigzag1}) completely
filled. There is a gap (which could be small) between the top of the
filled lower bulk band and the empty upper domain wall band
Eq.~(\ref{zigzag2}).  When they do overlap (as depicted in Fig.~2), some
of the upper domain wall band Eq.~(\ref{zigzag2}) will fill before all of
the negative energy bulk states are filled -- {\it the domain wall
will borrow some electrons from the bulk}.  It will then have a finite
charge density and a partially filled upper band which will behave
like a one-dimensional metal, even in neutral graphene.

The armchair domain wall is depicted in Fig.~1(c).  It is oriented
along the $y$-axis.  It has the same gapped bulk branches
Eq.~(\ref{bulk}) as the zig-zag wall. We look for wave-functions which
decay exponentially in distance $|A_x|,|B_x|$ from the wall.  They are
superpositions of two plane-waves propagating along the wall with
wave-vectors $k_y$ and $k_y+2\pi/3$.  (This corresponds to mixing of
the graphene valleys.)  The spectrum has four bands depicted in
Fig.~2. It is
$E=\pm \sqrt{ t^2\sin^2\frac{3a}{2}k_y + \mu^2 K(k_y)}$
 where
$K\cdot(1-e^{-\sqrt{3}ak_x^{(1)}})(1-e^{-\sqrt{3}ak_x^{(2)}})
 = \left(1+e^{-\frac{\sqrt{3}a}{2}(k_x^{(1)}+k_x^{(2)})}
\right)^2
$
 $k_x^{(1)}$ and $k_x^{(2)}$
must be determined by solving two equations:
$K-1=\frac{t^2}{\mu^2}\left(\cosh\frac{\sqrt{3}a}{2}k_x^{(1)}
+\cosh\frac{\sqrt{3}a}{2}k_x^{(2)}\right)^2$
and
$\cosh\frac{\sqrt{3}a}{2}k_x^{(2)}=\cosh\frac{\sqrt{3}a}{2}k_x^{(1)}+
\cos\frac{{3}a}{2}k_y ~,~ |k_y|\leq\frac{\pi}{3a}$.
We can find explicit solutions in the large and small ${\mu}/{t}$
limits.  When ${\mu}/{t}$ is large the spectrum is concentrated at
four values
\begin{equation}\mu>>t:~~E\approx\left\{ \matrix{\pm(\mu+t+\ldots) \cr
\pm(\mu-t+\ldots) \cr }\right.
\label{largemu}
\end{equation}
Two of these are inside the bulk spectrum and two are in the gap. They
agree with what we would expect when $\mu$ is large, where there are two energy
levels, $-\mu$ and $+\mu$ corresponding to electrons sitting on
the black or white dots, respectively in Fig.~1(c).  Then, the leading
effect of turning on a small $t$ is that the pairs of adjacent
degenerate states that exist at the location of the domain wall are
split by tunneling.  Now, unlike for the zig-zag, there are degenerate
pairs with both zeroth order energies $+\mu$ and $-\mu$.  The
splitting produces four
domain wall energies in Eq.~(\ref{largemu}). Further corrections are of order ${t^2}/{\mu}$
which, when taken into account spread the four levels into four
bands. which then get wider as $t$ gets larger.

In the limit $\mu<<t$ we also find four bands,
\begin{eqnarray}
\mu<<t:~~E\approx \left\{
\matrix{ \pm\sqrt{
\mu^2 +4t^2\left(1+\cos\frac{3a}{2}k_y\right)} \cr
\pm\sqrt{t^2\sin^2\frac{3a}{2}k_y +\frac{\mu^4}{t^2}\frac{4}{(4-
\cos^2\frac{3a}{2}k_y)^2} } \cr
}\right.
\end{eqnarray}
As depicted in Fig.~2, the upper and lower band are entirely within
the upper and lower bulk bands.  The middle two overlap the bulk
energy gap and themselves have a gap which is much smaller than the
bulk gap, $\Delta E={4\mu^2}/{3t}<<2\mu$.  For a typical small
$\mu\sim25mev$ and $t\sim2.7ev$, the gap in the mid-band states is
tiny, less than $1mev$.  This existence of a gap in the spectrum of
states bound to the domain wall is compatible with the continuum
analysis since the gap is vanishingly small in the continuum limit,
scaling to zero with the lattice spacing $a$, so it is not visible to
the continuum Dirac Hamiltonian.  Intuition for the gap in the
armchair spectrum can also be gained by studying Fig.~1(c).  Because of
the alternating pattern of pairs of black and pairs of white dots as
one follows the domain wall, the translation symmetry along the domain
wall is by two lattice spacings, rather than one.  This reduced
translation symmetry will gap the domain wall spectrum, analogous to
gapping created by a Peierls instability.  What is surprising here is
that the gap is so small.

In summary, we have shown that the simplest domain walls in gapped
graphene can have interesting electronic properties.  A partially
filled domain wall band will behave like a one-dimensional metal.
The continuum analysis suggests that similar behavior can be
expected for other types of domain walls, such as those arising from
reversing the pattern of a lattice distortion.  Analysis of the
details of the spectrum in those cases is left to future work. They
could also occur in other materials which have a Dirac spectrum,
such as the hypothetical flux phases of a square lattice where
vortices have recently been discussed \cite{franz} and where domain
walls, some of which would support zero modes should exist.

G.W.S.~thanks Peter Orland for discussions and the Institute Galileo
Galilei where part of the work was done;
we also thank Jun Liang Song for help with Fig.~2. This work was supported in
part by NSERC (Canada), Canadian Institute for Advanced research and the A. P. Sloan
Foundation.


\begin{thebibliography}{99}


\bibitem{Semenoff:1984dq}
  G.~W.~Semenoff,
  Phys.~Rev.~Lett.~{\bf 53}, 2449 (1984).

\bibitem{Niemi}
  A.~J.~Niemi, G.~W.~Semenoff,
  Phys.~Rev.~Lett.~{\bf 51}, 2077 (1983).

\bibitem{jackiw}R.~Jackiw, Phys.~Rev.~{\bf D29}, 2375 (1984)

\bibitem{novoselov:2004}
K.~S.~Novoselov et.~al., Science, Vol 306, 666 (2004)

\bibitem{Novoselov:2005kj}
  K.~S.~Novoselov {\it et al.},
  Nature {\bf 438}, 197 (2005)

\bibitem{geim}A.~K.~Geim, K.~S.~Novoselov,
Nat. Mater. {\bf 6}, 183 (2007).

\bibitem{kats}M. I. Katsnelson, Materials Today {\bf 10}, 20 (2007).


\bibitem{mindthegap}
K.~Novoselov, Nature Materials {\bf 6}, 720 - 721 (2007)


\bibitem{chamon:2007fr}
C.-Y.~Hou, C.~Chamon, C.~Mudry, Phys.~Rev.~Lett.~{\bf 98}, 186809 (2007)

\bibitem{nanotubes}C.~Chamon, Phys.~Rev.~{\bf B62}, 2806 (2000).

\bibitem{Novoselov:PNAS}
K.~S.~Novoselov et.~al., PNAS {\bf 102}, 30, 10451 (2005).

\bibitem{bn}
G.~Giovannetti, P.~A.~Khomyakov, G.~Brocks, P.~J.~Kelly, J.~van den Brink,
Phys.~Rev.~{\bf B76}, 073103 (2007).

\bibitem{lanzara}
S.Y.~Zhou, et.al.
Nature Materials {\bf 6}, 770 (2007).










 \bibitem{halleffect} Y.~Zhang, Y.-W.~Tan, H.~L.~Stormer, P.~ Kim,
Nature {\bf 438}, 201(2005).

\bibitem{Gusynin:2005pk} 
V.~P.~Gusynin, S.~G.~Sharapov,
  Phys.~Rev.~Lett.~{\bf 95}, 146801(2005):
N.~M.~R.~Peres, F.~Guinea, A.~H.~Castro Neto,
Phys.~Rev.~{\bf B73}, 125411(2006).



\bibitem{JackiwPi}
R.~Jackiw, S.-Y.~Pi, Phys.~Rev.~Lett.~{\bf 98}, 266402 (2007).

\bibitem{Chamon:2007pf}
C.~Chamon, C.~Y.~Hou, R.~Jackiw, C.~Mudry, S.~Y.~Pi, A.~P.~Schnyder,
  arXiv:0707.0293.

\bibitem{Chamon:2007hx}
C.~Chamon, C.~Y.~Hou, R.~Jackiw, C.~Mudry, S.~Y.~Pi, G.~Semenoff,
  arXiv:0712.2439 [hep-th].

\bibitem{Pachos}
J.~Pachos, M.~Stone, K.~Temme
Phys.~Rev.~Lett.~{\bf 100}, 156806 (2008)



\bibitem{polyacetylene1}
W.~P.~Su, J.~R.~Schrieffer, A.~Heeger, Phys.~Rev.~Lett.~{\bf 42}, 1698 (1979).

\bibitem{polyacetylene2}
R.~Jackiw, J.~R.~Schrieffer, Nucl,~Phys.~{\bf B190}, 253 (1981).

\bibitem{polyacetylene3}
A.~Niemi, G.~Semenoff, Phys.~Rep.~{\bf 135}, 99 (1986).



\bibitem{wilczek}
P.~Ghaemi, F.~Wilczek,  arXiv:0709.2626.

\bibitem{footnote} To formulate a general critereon for the existence
of the mid-gap band, assume that the Dirac operator has the form
$...+i\gamma^x\frac{\partial}{\partial x}+M(x)$ where $M$ is a mass
matrix.  Zero modes will exist when the two matrices
$i\gamma^xM(\infty)$ and $i\gamma^xM(-\infty)$ have simultaneous
eigenvalues whose real parts are positive and negative,
respectively. Then, if $s$ is the eigen-spinor, $\exp(-i\int _0^x
dx'\gamma^xM(x'))s$ is a normalizable wave-function.



\bibitem{wallace}
P.~R.~Wallace, Phys.~Rev.~{\bf 71}, 622 (1947).





\bibitem{franz}B.~Seradjeh, M.~Franz, arXiv:0709.4258 (2007).

\end{thebibliography}
\end{document}